\documentclass[final,3p,times,twocolumn]{elsarticle}
 \biboptions{comma,sort&compress}
\usepackage{here}
 \usepackage{graphicx}
  \usepackage{epsfig}

\def\nin{\noindent}
\def\beq{\begin{equation}}
\def\eeq{\end{equation}}
\def\bea{\begin{eqnarray}}
\def\eea{\end{eqnarray}}

\def\tr{{\rm tr}}

\newcommand{\MeV}{\,{\rm MeV}}

\journal{Nuc. Phys. (Proc. Suppl.)}

\begin{document}

\begin{frontmatter}



\title{The hadron resonance gas model: thermodynamics of QCD and Polyakov loop}

 \author[label1]{E.~Meg\'{\i}as\corref{cor1}}
  \address[label1]{Grup de F\'{\i}sica Te\`orica and IFAE, Departament de F\'{\i}sica, Universitat Aut\`onoma de Barcelona, Bellaterra E-08193 Barcelona, Spain}
\cortext[cor1]{Speaker}
\ead{emegias@ifae.es}

 \author[label2]{E.~Ruiz Arriola} 
   \address[label2]{Departamento de F{\'\i}sica At\'omica, Molecular y Nuclear and Instituto Carlos I de F{\'\i}sica Te\'orica y Computacional, Universidad de Granada, E-18071 Granada, Spain.}
\ead{earriola@ugr.es}

\author[label2]{L.L.~Salcedo}
\ead{salcedo@ugr.es}

\begin{abstract}
\noindent
We study the hadron resonance gas model and describe the equation of state of QCD and the vacuum expectation value of the Polyakov loop in the confined phase, in terms of hadronic states with light quarks in the first case, and with exactly one heavy quark in the second case. Comparison with lattice simulations is made.
\end{abstract}


\end{frontmatter}


\section{Introduction}
\label{sec:introduction}
\nin
Two symmetries are relevant in the study of the thermodynamics of QCD~\cite{Fukushima:2011jc}. On the one hand the chiral symmetry is broken at low temperatures, while it becomes restored above a certain temperature~$T_\chi$. In the limit of massless quarks, the order parameter for this transition is the quark condensate~$\langle \bar q q\rangle$. On the other hand, the center symmetry~$\mathbb{Z}(N_c)$ of the gauge group SU($N_c$) is broken at high temperatures and it is restored below~$T_c$, known as deconfinement temperature. This symmetry controls the confinement/deconfinement of color charges~\cite{Svetitsky:1985ye}. In the limit of infinitely heavy quark masses (gluodynamics), the order parameter for this transition is the Polyakov loop~$L_T = \langle \tr_c {\sf P} e^{i \int_0^{1/T} \! A_0\, dx_0 } \rangle $, where $A_0$ is the gluon field and ${\sf P}$ denotes path ordering. It is accepted nowadays that both transition temperatures are very close to each other, $T_\chi \approx T_c$, at least at zero chemical potential~\cite{Coleman:1980mx,Meisinger:1995ih,Sakai:2010rp,Braun:2011fw}. 

While in the deconfined phase of QCD the quarks and gluons are liberated to form a plasma, in the confined/chiral symmetry broken phase the relevant degrees of freedom are bound states of quarks and gluons, i.e. hadrons and possibly glueballs. This means that it should be expected that physical quantities in this phase admit a representation in terms of hadronic states. This is the idea of the hadron resonance gas model (HRGM) which describes the equation of state of QCD in terms of a free gas of hadrons~\cite{Hagedorn:1984hz,Yukalov:1997jk,Agasian:2001bj,Tawfik:2004sw, Megias:2009mp,Huovinen:2009yb,Borsanyi:2010cj,Bazavov:2011nk},
\begin{eqnarray}
&&\!\!\!\!\!\!\!\!\!\!\!\!\!\!\!\!\!\!\!\frac{1}{V}\log Z  \label{eq:hrgm} \\
&&\!\!\!\!\!\!\!\!\!\!\!\!\!\!\!= -\int \frac{d^3 p}{(2\pi)^3} \sum_\alpha \zeta_\alpha g_\alpha 
\log \left( 1 - \zeta_\alpha e^{-\sqrt{p^2+M_\alpha^2}/T} \right) \,, \nonumber
\end{eqnarray}
with $g_\alpha$ the degeneracy factor, $\zeta_\alpha=\pm 1$ for bosons and
fermions respectively, and $M_\alpha$ the hadron mass. It has been presented in~\cite{Megias:2012kb,Arriola:2012wd} a similar model to describe the Polyakov loop in terms of hadronic resonances with exactly one heavy quark. In this communication we will ellaborate on these models, and perform a comparison with recent lattice simulations.

\section{The hadron resonance gas model and the Polyakov loop}
\label{sec:hrgm}
\nin
An effective approach to the physics of the phase transition is provided by chiral quark models coupled to gluon fields in the form of a Polyakov loop~\cite{Meisinger:1995ih,Fukushima:2003fw,Megias:2004hj,Megias:2006bn,Ratti:2005jh,Schaefer:2007pw,Contrera:2007wu,Lourenco:2012dx}. Most of these works remain within a mean field approximation and assume a {\it global} Polyakov loop. Based on QCD arguments we have shown in~\cite{Megias:2012kb} that a hadronic representation of the Polyakov loop is given by
\begin{equation}
L_T = \langle \tr_c {\sf P} e^{i \int_0^{1/T} \! A_0\, dx_0 } \rangle  \approx \frac{1}{2} \sum_\alpha g_{h\alpha} e^{-\Delta_{h\alpha} / T} \,, \label{eq:hrgm_PL}
\end{equation}
where $g_{h\alpha}$ are the degeneracies and $\Delta_{h\alpha} = M_{h\alpha} -
m_h$ are the masses of hadrons with exactly one heavy quark (the mass
of the heavy quark itself $m_h$ being subtracted). We have shown in~\cite{Arriola:2012wd} that Eq.~(\ref{eq:hrgm_PL}) is fulfilled in chiral quark models coupled to the Polyakov, when one goes beyond mean field and advocate the local and quantum nature of the Polyakov loop. The need of these corrections was already stressed in~\cite{Megias:2004hj,Megias:2006bn,Megias:2005qf,Megias:2006df}.

A natural step is to check to what extent the hadronic sum rule is
fulfilled by experimental states compiled in the
PDG~\cite{Nakamura:2010zzi}. The relation~(\ref{eq:hrgm_PL}) follows
in the limit~$m_h \to \infty$, but the heavy quarks in nature have a
finite mass. Hadrons with a bottom quark would be optimal, due to the
large quark mass compared to $\Lambda_{\rm QCD}$, but the available
data are scarce, so we turn to charmed hadrons. Specifically, we
consider the lowest-lying single-charmed mesons and baryons with $u$,
$d$, and $s$ as the dynamical flavors, with quarks in relative
$s$-wave inside the hadron. For mesons, these are usually identified
with the states (spin-isospin multiplets) $\bar{D}$, $\bar{D}_s$,
$\bar{D}^*(2010)$ and $\bar{D}_s^*$, and for baryons, with
$\Lambda_c$, $\Sigma_c(2455)$, $\Xi_c$, $\Xi_c^\prime$, $\Omega_c$,
$\Sigma_c(2520)$, $\Xi_c(2645)$, and $\Omega(2770)$. A total of~12
meson states and 42 baryon states.

\begin{figure}[ttt]
\begin{center}
\centerline{\includegraphics[width=7.3cm,height=5cm]{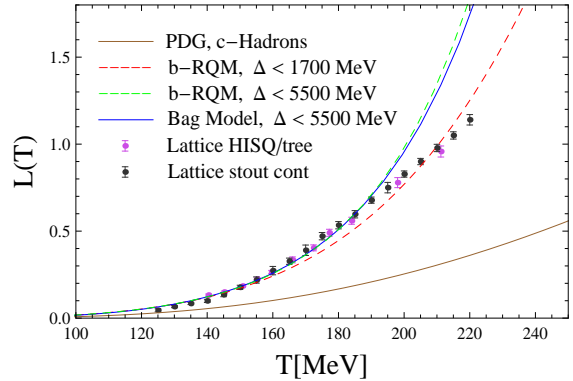}}
\end{center}
\vspace{-0.8cm}
\caption{Polyakov loop as a function of temperature (in MeV). Lattice data from~\cite{Bazavov:2011nk} for the HISQ/tree action with $N_\sigma^3\times N_\tau = 32^3\times 12$, and from~\cite{Borsanyi:2010bp} for the continuum extrapolated stout result. The solid (brown) line shows the sum rule~(\ref{eq:hrgm_PL}) saturated with the lowest-lying charmed hadrons from PDG~\cite{Nakamura:2010zzi}. Dashed lines correspond to the result using the RQM spectrum with one $b$ quark and a cut-off $\Delta < 1700 \,\MeV$ (red line), and $\Delta < 5500 \, \MeV$ (green line). The result from the MIT bag model with cut-off $\Delta < 5500\, \MeV$ is shown as a solid (blue) line~\cite{Megias:2012kb}. The results from PDG and RQM have been multiplied by a factor $L(T)\to e^{C/T} L(T)$, with $C=25\,\MeV$, which comes from an arbitrariness in the renormalization.}
\label{fig:rqmbag}
\end{figure}

The plot in Fig.~\ref{fig:rqmbag} shows that the lowest-lying states
fall short to saturate the sum rule, regardless of the choice of mass
of the charmed quark, $m_c$. This is not surprising as any model
predicts many excited states on top of the lowest-lying ones, as is
also the case for light-quark hadrons. Adding more states from the PDG
does not seem practical due to the fragmentary information
available. Instead we turn in what follows to hadronic models. The aim
is not so much to have a detailed description of the various states
but to give a sufficiently good overall description of the whole
spectrum.

\section{The relativized quark model: trace anomaly and Polyakov loop}
\label{sec:RQM}
\nin

Several models in the past have been proposed to describe the hadron spectrum and give a prediction for excited states not yet included in the PDG. One of them is the relativized quark model (RQM)~\cite{Godfrey:1985xj,Capstick:1986bm}; it is a soft QCD model based on a one-gluon exchange at short distances and a phenomenological implementation of confinement by a flavor-independent Lorentz-scalar interaction. We will use the spectrum predicted with this model to saturate the sum rules.

\subsection{Thermodynamics of QCD}
\label{subsec:thermodynamics}
\nin

\begin{figure}[ttt]
\begin{center}
\centerline{\includegraphics[width=7.3cm,height=5cm]{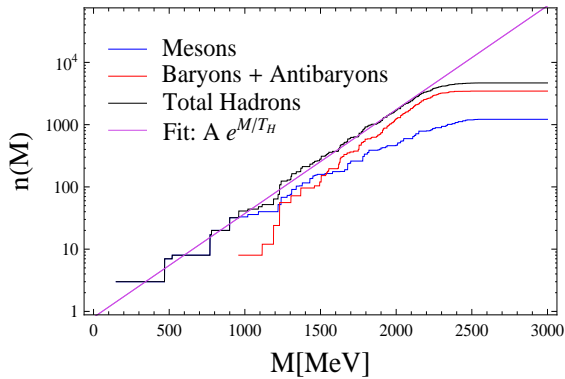}}
\end{center}
\vspace{-0.8cm}
\caption{Cumulative number $n$ as a function of the hadron mass $M$ (in $\MeV$), for meson spectrum (blue line), baryon spectrum (red line) and total number of hadrons (black line) with $u$, $d$ and $s$ quarks, computed in the relativized quark model of Refs.~\cite{Godfrey:1985xj,Capstick:1986bm}. We also plot a fit of the total cumulative number using Eq.~(\ref{eq:nHagedorn}) (magenta line).}
\label{fig:cumulative}
\end{figure}

From the standard thermodynamic relations, the free energy, the pressure, and the energy density are given by
\begin{equation}
F = - pV = -T \log Z \,, \quad \epsilon = \frac{E}{V} = \frac{T^2}{V} \frac{\partial \log Z}{\partial T} \,,
\end{equation}
as well as the relation for the trace anomaly
\begin{equation}
{\cal A}(T) \equiv \frac{\epsilon - 3p}{T^4} = T\frac{\partial}{\partial T}\left( \frac{p}{T^4} \right) \,.
\end{equation}
The Hagedorn formula for the trace anomaly follows from Eq.~(\ref{eq:hrgm}) and the relations given above. It writes 
\begin{eqnarray}
&&\!\!\!\!\!\!\!\!\!\!\!\!\!\!\!\!\!\!\!\!\!\!\frac{\epsilon - 3p}{T^4} = \sum_{k=1}^\infty \int dM \left( \frac{\partial n_m(M)}{\partial M} + (-1)^{k+1}\frac{\partial n_b(M)}{\partial M} \right) \nonumber \\
&&\qquad\times \frac{1}{2 k \pi^2} \left( \frac{M}{T} \right)^3 K_1\left(k \frac{M}{T}\right) \,,\label{eq:traceanomalyHagedorn}
\end{eqnarray} 
where $K_1(z)$ refers to the first order Bessel function. $n_m$ and $n_b$ are the cumulative numbers of mesons and baryons (including antibaryons), defined as
\begin{equation}
n(M) = \sum_\alpha g_\alpha \Theta( M - M_\alpha ) \,.
\end{equation}
 $\Theta$ is the step function. $n(M)$ represents the number of
hadrons with mass less than~$M$. Hagedorn proposed that the cumulative
number of hadrons in QCD is approximately given by
\begin{equation}
n(M) = A \, e^{M/T_H} \,, \label{eq:nHagedorn}
\end{equation}
where $A$ is a constant and $T_H$ is the so called Hagedorn
temperature. We show in Fig.~\ref{fig:cumulative} the cumulative
number of hadrons with $u$, $d$ and $s$ quarks computed in the
RQM. The curves increase up to a cutoff $M\approx 2300\, \MeV$, above
which they become approximately flat. The total cumulative number can
be approximated to the form of eq.~(\ref{eq:nHagedorn}), with $A =
0.80$, $T_H = 260\, \MeV$ and $\chi^2/\textrm{dof}=0.031$, in the
regime $500\,\MeV < M < 2300\,\MeV$.

\begin{figure}[ttt]
\begin{center}
\centerline{\includegraphics[width=7.3cm,height=5cm]{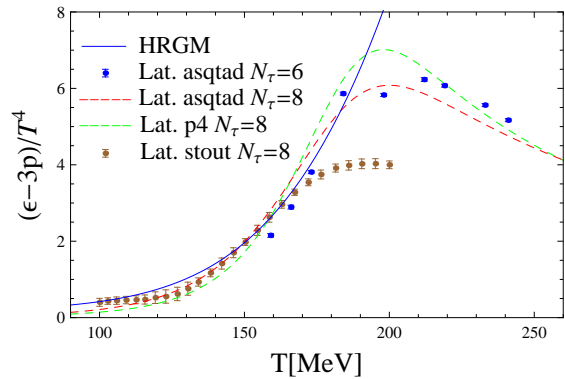}}
\end{center}
\vspace{-0.8cm}
\caption{Trace anomaly $(\epsilon - 3p)/T^4$ as a function of temperature (in MeV). Blue points are lattice data with the asqtad action and $N_\tau = 6$ from~\cite{Bazavov:2009zn}. Dashed lines correspond to a parameterization given in that reference of the lattice data with p4 and asqtad actions for $N_\tau = 8$. Brown points are lattice data with the stout action $N_\tau=8$ from~\cite{Borsanyi:2010bp}. The result from the HRGM Eq.~(\ref{eq:traceanomalyHagedorn}), computed with the RQM spectrum with $u$, $d$ and $s$ quarks from Refs.~\cite{Godfrey:1985xj,Capstick:1986bm}, is  shown as a solid (blue) line. A temperature shift of $T_0 = 15\,\MeV$ to lower temperatures is introduced in the lattice data from~\cite{Bazavov:2009zn}.}
\label{fig:traceanomaly}
\end{figure}

As a cross-check of the RQM we can use the spectrum obtained with this
model to compute the trace anomaly using the HRGM given by
Eq.~(\ref{eq:traceanomalyHagedorn}). The result and its comparison
with lattice data is shown in Fig.~\ref{fig:traceanomaly}. The RQM
gives a good description of the trace anomaly for $T <
180\,\MeV$.~\footnote{In order to get a good agreement with lattice
  data from~\cite{Bazavov:2009zn}, a temperature shift of $T_0 \approx
  15\,\MeV$ is required; ${\cal A}_{\textrm{\tiny HRGM}} (T) = {\cal
    A}_{\textrm{\tiny Lattice}}(T+T_0)$. The need of this shift in
  lattice data was also observed in~\cite{Bazavov:2009zn}, and it has
  been attributed to systematic errors in lattice coming from
  extrapolations to the physical light quark masses and the continuum
  limit.}

\subsection{Polyakov loop}
\label{subsec:Polyakov_loop}
\nin

The next step is to use the spectrum of hadrons with one heavy quark
at rest predicted by the RQM to compute the Polyakov loop using the
HRGM of Eq.~(\ref{eq:hrgm_PL}). The total number of hadron states
computed in \cite{Godfrey:1985xj,Capstick:1986bm} with one $c$ quark
is $117$ for mesons and $660$ for baryons, corresponding to a maximum
value of $\Delta=M-m_c$ about $1700\MeV$. For hadrons with one $b$
quark, $87$ mesonic and $776$ baryonic states, with a similar upper
bound for~$\Delta$. In these papers there are some missing states
corresponding to baryons of the type $csu$ and $bsu$, in particular
$\Xi_{c,b}$, $\Xi^\prime_{c,b}$ and $\Omega_{c,b}$ baryons. In order
to give a prediction for these states we have used the equal spacing
rule~\cite{Savage:1995dw}, which is based on the approximate relations
\begin{eqnarray}
&&M_{\Xi_c} - M_{\Lambda_c} \approx m_s \,, \qquad M_{\Xi_c^\prime} - M_{\Sigma_c} \approx m_s \,, \nonumber \\
&& M_{\Omega_c} - M_{\Sigma_c} \approx 2m_s \,.
\end{eqnarray}
They follow from the replacement of a light quark ($u$ or $d$ quark) in $\Lambda_c$ by a $s$ quark to get $\Xi_c$, and assuming $m_s \gg m_u \,, m_d$. The same replacement in $\Sigma_c$ with one $s$ quark allows to get $\Xi_c^\prime$,  and with two $s$ quarks to get~$\Omega_c$. These relations are valid also for baryons with one $b$ quark, and they preserve the degeneracy of states. After including these missing states, the total number of baryons states we consider with one $c$ quark is $1470$, and with one $b$ quark is $1740$.  We use an $s$ quark mass of $109\,\MeV$ (extracted from the lowest-lying hadrons). 

\begin{figure}[ttt]
\begin{center}
\centerline{\includegraphics[width=7.3cm,height=5cm]{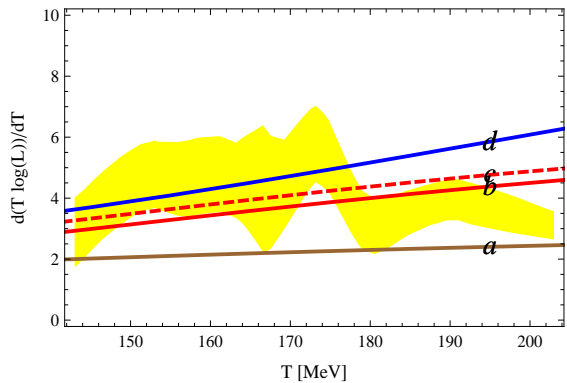}}
\end{center}
\vspace{-0.8cm}
\caption{Comparison of $\frac{d}{dT}(T\log(L(T)))$ (yellow strip) with
  $\frac{d}{dT}(T\log(\frac{1}{2}\sum_\alpha g_{h\alpha}
  e^{-\Delta_{h\alpha}/T}))$ from hadronic states (mesons plus
  baryons): lowest-lying hadrons from PDG (solid brown line, label
  $a$), RQM states from \cite{Godfrey:1985xj,Capstick:1986bm} with
  quark $c$ (solid red line, label $b$), and with quark $b$ (dashed
  red line, label $c$), and MIT bag model estimate including states up
  to $\Delta=5500\MeV$ (solid blue line, label $d$)~\cite{Megias:2012kb}.}
\label{fig:3}
\end{figure}

The prediction based on these hadronic states is displayed in
Fig.~\ref{fig:rqmbag}. The RQM result is closer to the lattice data
than the naive estimation from the lowest-lying hadrons in PDG, but
still tend to stay below it, a consequence of the truncation of states
to $\Delta<1700\MeV$. In order to remove any ambiguities coming from
the renormalization prescription used for the Polyakov loop, we plot
in Fig.~\ref{fig:3} the derivative of $T\log(L(T))$ with respect to
$T$.~\footnote{The yellow strip in Fig.~\ref{fig:3} is the uncertainty
  from a combined analysis of lattice data from continuum extrapolated
  stout~\cite{Borsanyi:2010bp}, HISQ/tree action $N_\tau = 12$ scale
  sets $r_1$ and $f_k$, and asqtad scale $f_k$~\cite{Bazavov:2011nk}.}
It is noteworthy that the bottom sum rule gives a better value than
the charm one, as it would be expected due to the larger mass of the
$b$ quark.

In order to avoid the cut-off problem in the spectrum, we have assumed a raising of cumulative numbers of the form $n(\Delta) = A \, \Delta^{\alpha}$ in the regime $1700\,\MeV < \Delta < 5500 \,\MeV$, with values $A=5.92\cdot 10^{-9}$, $\alpha=3.16$ for $b$-mesons, and $A=1.76\cdot 10^{-24}$, $\alpha=8.23$ for $b$-baryons.~\footnote{These numbers follow from a fit of $n(\Delta)$ below $1700\,\MeV$.} The analytic formula to be applied in this regime is
\begin{equation}
L(T) = \frac{1}{2} \int d\Delta \frac{\partial n(\Delta)}{\partial \Delta} e^{-\Delta/T} \,,
\end{equation}
which gives a contribution to be added to the numerical one below $1700\,\MeV$.  The result is displayed in Fig.~\ref{fig:rqmbag} with a dashed (green) line. The estimated effect of adding the states up to $\Delta = 5500 \,\MeV$ in the RQM leads to a result quite consistent with the one obtained with the MIT bag model discussed in~\cite{Megias:2012kb}, with the same cut-off. We have also checked that this result is numerically indistinguishable from that obtained by extending the analytic raising up to $\Delta\to\infty$ (no cut-off).  

\section{Conclusions}
\label{sec:conclusions}
\nin
The hadron resonance gas model was proposed as a simple model to describe the confined phase of QCD in terms of a free gas of hadronic states (mesons and baryons). Using the hadronic spectrum obtained by the relativized quark model of Ref.~\cite{Godfrey:1985xj,Capstick:1986bm}, we get a good description of lattice data for the trace anomaly up to $T = 180 \,\MeV$. A different version of the hadron resonance gas model has been proposed in~\cite{Megias:2012kb,Arriola:2012wd} to describe the vacuum expectation value of the Polyakov loop in terms of hadronic states with exactly one heavy quark at rest and several dynamical quarks. The lowest-lying charmed mesons and baryons included in the PDG gives for the Polyakov loop a value well below lattice data, and this suggest the need for inclusion of more states. When using the spectrum predicted by the relativized quark model and the MIT bag models up to a cut-off $\Delta = 5500\,\MeV$, we get a good description of lattice data for the Polyakov loop in the confined phase.

This work opens the possibility of a Polyakov loop spectroscopy, i.e. using the Polyakov loop in fundamental and higher representations to deduce multiquark states, gluelumps, etc, containing one or several heavy quark states.

\section*{Acknowledgements}
\nin
This work has been supported by DGI (FIS2011-24149 and FPA2011-25948)
and Junta de Andaluc{\'\i}a grant FQM-225. The research of
E.~Meg\'{\i}as is supported by the Juan de la Cierva Program of the
Spanish MICINN.









\end{document}